\documentclass[iop]{emulateapj}
\usepackage{graphicx,morefloats}
\usepackage{natbib,color,epsfig}
 \pdfoutput=1           
\newcommand{\kms}{km~s$^{-1}$}
\newcommand{\cc}{cm$^{-3}$}
\newcommand{\DNs}{DN~s$^{-1}$}

\newcommand{\SiIV}{Si~IV}

\newcommand{\OVI}{O~VI}
\newcommand{\NeVIII}{Ne~VIII}

\begin{document}
%\doublespace

\title{\bf IRIS \SiIV\ LINE PROFILES: AN INDICATION FOR THE PLASMOID INSTABILITY DURING SMALL-SCALE MAGNETIC RECONNECTION ON THE SUN}
\author{D.E. INNES\altaffilmark{1,2}, L.-J. GUO\altaffilmark{1,2}, Y.-M. HUANG\altaffilmark{2,3}, 
A. BHATTACHARJEE\altaffilmark{2,3,4}}
%\offprints{D.E. Innes (Corresponding author: innes@mps.mpg.de)}
\affil{$^1$Max Planck Institute for Solar System Research, D-37077 G\"{o}ttingen, Germany}
\affil{$^2$Max Planck/Princeton Center for Plasma Physics, Princeton, NJ 08540, USA}
\affil{$^3$Department of
Astrophysical Sciences and Princeton Plasma Physics Laboratory, Princeton
University, Princeton, NJ 08540, USA}
\affil{$^4$Center for Heliophysics, Princeton University, Princeton, NJ 08540, USA}
\email{innes@mps.mpg.de} 
%\date{Received; accepted}

%\title{Cascades of magnetic islands  during fast magnetic reconnection on the Sun}
%\author{D.E. Innes, L.-J. Guo, Y.-M. Huang, A. Bhattacharjee}

%*****************************************************************************
\begin{abstract}
Our understanding of the process of fast reconnection
has undergone a dramatic change in the last 10 years driven, in part, by the
availability of high-resolution numerical simulations that have consistently
demonstrated the break-up of current sheets into magnetic islands,
with  reconnection rates that become independent of Lundquist number, challenging the belief that  fast magnetic reconnection in flares proceeds via
the Petschek mechanism that invokes pairs of slow-mode shocks
connected to a compact diffusion region. 
%The fundamentally different
%current sheet configurations have repercussions on particle accleration.
%in the Earth's magnetosphere, at the Sun, and in astrophysical sources.
%Space plasma conditions cannot be reproduced in the laboratory, so identifying the
%basic reconnection mechanism has to be done in space.
The
 reconnection sites are too small to be resolved with
images but these reconnection mechanisms, Petschek and the plasmoid instability,  have reconnection sites with very different density and
velocity structures and so can be distinguished by high-resolution
line-profiles observations.
Using IRIS spectroscopic observations we obtain  a survey of typical 
line profiles produced by small-scale events thought to be reconnection sites on
the Sun.
% and compare them with theoretical line profiles of reconnecting current sheets to
%determine whether reconnection via the plasmoid instabilty can explain the observed profiles.
%We use high-resolution sit-and-stare spectral observations of the \SiIV\ line,
%obtained by the IRIS spectrometer, to identify sites of acceleration,  and follow the development of line profiles. 
%IRIS high-resolution 
Slit-jaw images are used to investigate the plasma heating and re-configuration at the sites.
A sample of 15 events from two active regions is presented. The line profiles are complex with bright cores and broad wings extending to over 300~\kms. The profiles can 
be reproduced with  the multiple magnetic islands and acceleration sites that characterise the plasmoid instability but not by bi-directional jets that characterise the Petschek mechanism.
 This result suggests that if these small-scale events are reconnection sites, then fast reconnection proceeds via the plasmoid instability, rather than the Petschek mechanism during small-scale reconnection on the Sun.
%This result suggests 

\end{abstract}
\keywords{magnetic reconnection - Sun:activity  - Sun: transition region - Sun: UV radiation} 
%\titlerunning{Small-scale reconnection on the Sun}
%\authorrunning{Innes et al}

%\maketitle

%*****************************************************************************

%*****************************************************************************
%%%%%% Section 1 %%%%%%   %%%%%%%%%%%%%%%%%%%%%%%%%%%%%%%%%%%%%%%%%%%%%%%%%%%%
\section{Introduction}
Magnetic reconnection is the breaking and rejoining of magnetic connections.
In space plasmas, it is accompanied by a rapid conversion of magnetic energy
into plasma thermal and kinetic energy, and intense particle acceleration.
It is fundamental to our understanding of rapid energy
release  in space plasmas from the Earth's magnetotail \citep{Xiao06,Angelopoulos08,Nagai11} to very remote extragalactic jets \citep{Sironi14, Giannos13}.
At the Sun the consequences of reconnection are observed
directly in  coronal mass ejections (500-1000~Mm), flares (100-300~Mm), 
and ubiquitous small-scale jets (1~Mm). The Sun therefore
provides a  range of scales over which reconnection can be studied.
Diagnostics rely on images
of the high temperature ($\ge10$~MK) plasma emissions \citep{Masuda94,Aschwanden02,Savage10,Su13}, spectra
of  high velocity jets \citep{Detal91, IIAW97b, Winebarger02, Imada13}, radio emission \citep{Nish15}, and in-situ detections of energetic particles \citep{Klassen05,Klassen11}.

When reconnection occurs along large aspect-ratio current sheets, structures that appear to look  like current sheets with embedded plasmoids have been
seen, for example,  behind coronal mass ejections \citep{LinJ05,Lin07,Ciar08,LiuR10,Guo13}.
In large flares reconnection occurs at multiple sites simultaneously and
it is difficult to disentangle the dynamics of the reconnection process itself.
For this reason, we have investigated small-scale ($\sim$5000~km), short-lived ($1-5$~min) reconnection sites identified by broad non-Gaussian transition region line profiles, known as transition-region explosive events \citep{DBB89}. The line-profile evolution during events on the quiet Sun has been shown to be consistent with bi-directional flow \citep{Detal91, IIAW97b}, as predicted by reconnection models. It was thought that much of the acceleration and heating occurs, as  proposed by Petschek \citep{P64}, along  shocks attached to the reconnection region, so that  most of the plasma at the reconnection site moves with the Alfv\'en speed. In the realm of theory, there has been a clear consensus that the Petschek mechanism does not hold unless the resistivity of the plasma is enhanced locally at the X-point (see \citet{Biskamp00} and other references therein). Despite this theoretical reservation about the Petschek theory, there has been a general tendency to continue comparing observations with the theory. 
The Petschek model accounts for the high-velocity  component but  there is insufficient low velocity plasma in the diffusion region to reproduce the strong core brightening that is observed in many events along with the  wing enhancements \citep{Detal91,Innes01}.  Larger-scale simulations that incorporate surrounding chromospheric and coronal plasma show jets and brightening of the line core from low-velocity plasma outside the diffusion region  \citep{Heggland09, Ding11}, so the line core  emission has been attributed to heating of the background plasma. In these scenarios, the jet and background emission are spatially off-set which should result in spatial off-sets between the core and wing brightening.
%Therefore it was thought that the core brightening was due to heating in the surrounding plasma. 

The recently launched IRIS spectrometer has significantly higher spatial
and spectral resolution than previous solar ultraviolet (UV) spectrometers, thus allowing better segmentation
of the reconnection site flows. In addition, high spatial resolution (250~km)  co-temporal slit-jaw images show the surrounding transition region and chromospheric heating.
Here we focus on active-region events with broad non-Gaussian wings in the \SiIV\ line profile.
The \SiIV\ lines are formed over a narrow temperature range around $10^5$~K, are optically
thin, and change rapidly during  transition region  heating and acceleration processes. 

Extensive studies of similar line profiles seen in the quiet Sun  
 by HRTS \citep{BB83, DBB89, Detal91},
and SUMER have shown that they are consistently seen above flux cancellation sites at the boundary of coronal holes \citep{Madjarska12}, the junctions of supergranule cells \citep{Innes13}, and during flux emergence \citep{Detal91}. The line profiles usually show both core and wing enhancements with the wing often preceding the core brightening by about 1-2~min \citep{Innes01}. Sequences of small, fast rasters have shown that in many quiet-Sun events the red and blue wings brighten  simultaneously \citep{Ning04}, although they may be off-set along the slit \citep{DBB89,IIAW97b}.
The active region events sometimes have very complicated line profiles with superimposed narrow absorption lines from singly ionised, neutral and even molecular species \citep{Peter14,Schmit14}.

Although the association with reconnection is widely accepted the actual cause of the line brightening and broad wings is not well understood. Analysis of a quiet Sun explosive event observed by IRIS concluded that bi-directional jets were unlikely and suggested plasma ejection and retraction as a plausible scenario \citep{Huang14}. Here we present a variety of line profiles observed during explosive events and consider reconnection along a thin current sheet subject to the plasmoid instability as an explanation for the profiles seen. The observed line profiles are compared with synthetic profiles from 2D simulations of reconnecting current sheets.

\section{Observations}
For this study we use high-cadence sit-and-stare IRIS observations of the \SiIV\ 1402.77~\AA\ taken at full spectral (26~m\AA) and spatial (0.33\arcsec) resolution to investigate the time evolution of profiles during explosive events.
Context images of the reconnection sites
are provided by the IRIS slit-jaw camera at 1400~\AA, SDO/AIA \citep{Lemen12}  at 1600~\AA\ and 171~\AA, and SDO/HMI \citep{Scherrer12} magnetograms.
The 1400~\AA\ images  reveal sites of bright chromospheric continuum and \SiIV\ transition-region line emission.
They can be easily aligned to the AIA 1600~\AA\  images which are also dominated by chromospheric continuum and transition-region line emission. In the quiet-Sun regions most transition-region brightening coincides with 171~\AA\ brightening, but in active regions many of the transition region structures are only seen at wavelengths longward of the Lyman edge at 912~\AA\ because in active regions there is a significant amount of high lying neutral material that is optically thick at wavelengths less than 912~\AA\ due to hydrogen and helium photoionisation. Nevertheless there are enough common 1400 and 171~\AA\  features to obtain good image coalignment.
The HMI magnetograms were more difficult to coalign. First we coaligned the 1600~\AA\ to the 1700~\AA\ images. Then bright structures in the 1700~\AA\ images were matched with small flux concentrations in the magnetograms.

We show profiles from several sites in two active regions. One  was more active and brighter than the other and for this region a shorter exposure time was chosen. The exposure times were 2 and 4~s, and this resulted in 
time cadences of 3.5 and 5.6~s, allowing for the CCD read-out time (about 1.5~s). In  this study, we use sit-and-stare observations which show the evolution of line profiles at a single position. Although this has the disadvantage of measuring flows along a narrow line-of-sight through the event, rather than showing the flow geometry across the whole event, we obtain better resolution of the profile evolution than would be possible by rastering. The observed profiles can be readily compared to synthetic profiles from comparable slices of the simulated current sheets. 
%width of explosive events is $\sim6$\arcsec, so with a cadence of about 5~s and raster steps equal to the slit width, 0.33\arcsec, it would take about 90~s to cover the events which is  roughly the event lifetime. 
%In the future, we plan to use small, fast rasters to investigate the flow geometry.

\subsection{2014 April 15 observations of AR 12036}

\begin{figure}
%\centering
\includegraphics[width=7cm,clip=]{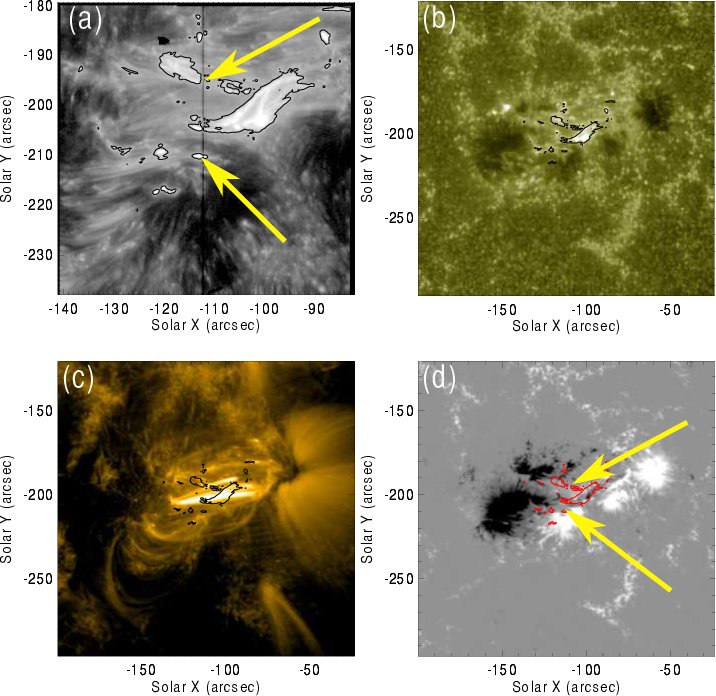}
\caption{AR 12036 on 2014 April 15 10:22:50~UT:
(a) IRIS 1400~\AA\ slit jaw; (b) AIA 1600~\AA;
(c) AIA 171~\AA; (d) HMI line-of-sight magnetogram. The black vertical line in
(a) is the IRIS spectrometer slit. All images are overplotted with the
same 1400~\AA\ contours at 150~\DNs. In (a) and (d) the lower yellow arrow points to the site of events 1 and 2, and the upper one to the site of 3, 4, and 5, marked in Fig~\ref{iw1}(b). } \label{over1}
\end{figure}

The first region is a recently emerged active region, AR12036 (Figure~\ref{over1}).
It  emerged three days earlier, on 2014 April 12, and
new flux was still emerging at the time of the observations around 10:00~UT on the 2014 April 15.  
There was a lot of small and large-scale activity in the region, including  a C4.4 flare at 09:53~UT, just 7~min before the observations presented here. The bright coronal loops seen in the 171~\AA\ image (Figure~\ref{over1}(c)) connecting positive and negative spots
appeared during this flare.
The main 1400~\AA\ brightening is below the loops and is visible at 1600~\AA\ but not in the  171~\AA\  or any of the other extreme ultraviolet (EUV) images. 
The spectrometer slit, shown as a dark vertical line in Figure~\ref{over1}(a), was oriented north-south along the neutral line in the north and across newly emerged positive 
flux in the south.
Both spectrometer and slit-jaw exposures were 2~s and image cadences were 3.5~s.

\begin{figure}
%\centering
\includegraphics[width=8.5cm,clip=]{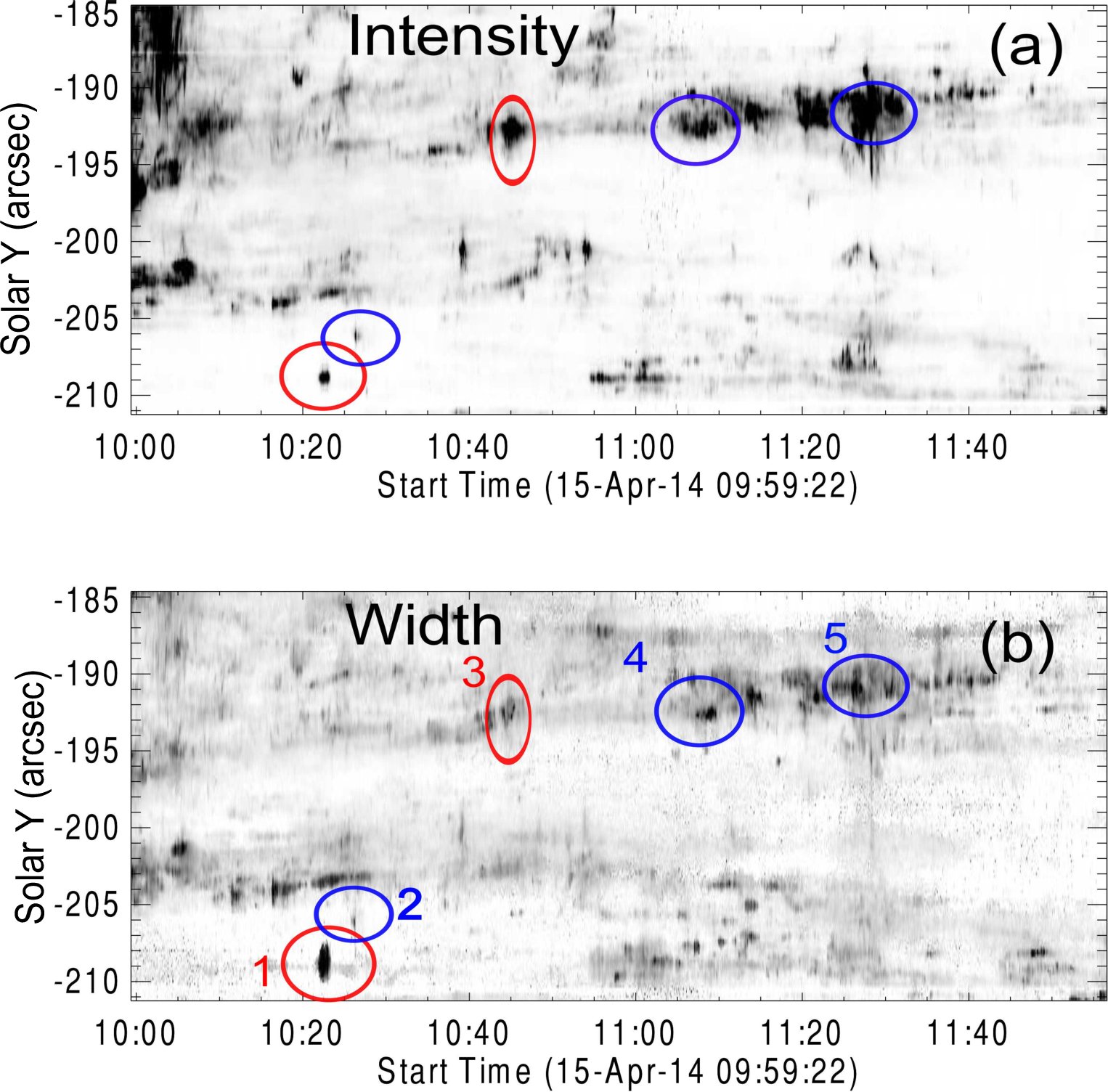}
\caption{\SiIV\ 1402~\AA\ (a) intensity and (b) width during two hours of sit-and-stare.
The maximum values are 2500~\DNs\ and 100~\kms, respectively. The events circled in red are discussed in the main paper. The profiles of those circled in blue are shown in the appendix.
} \label{iw1}
\end{figure}

\begin{figure}
%\centering
\includegraphics[width=8.5cm,clip=]{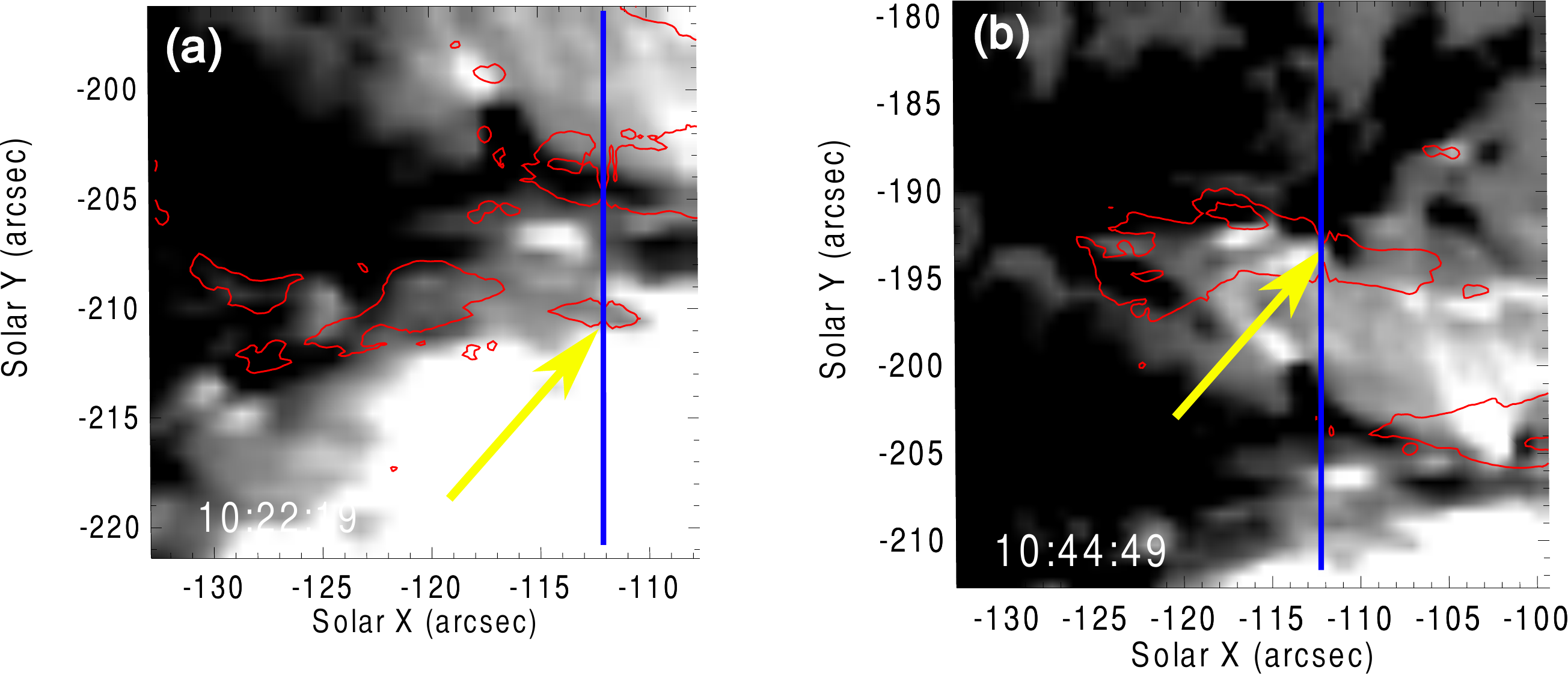}
\caption{Line-of-sight magnetic fields near the main sites of \SiIV\ explosive events: (a) the region around the events 1 and 2; (b)  the region around 3, 4 and 5. The blue vertical line indicates the position of the spectrometer slit. Contours are 1400~\AA\ at 100~\DNs.} \label{mfs_apr15}
\end{figure}

 In Figure~\ref{iw1} we show space-time maps of the \SiIV\  1402~\AA\ line intensity and width along the northern half of the slit where most the broad  \SiIV\  profiles were seen. Single-Gaussian line-profile fits have been used. The broadest profiles came mostly from sites just north and south of the bright 171~\AA\ loops, where the slit crossed the edge of a region of strong magnetic field, marked with yellow arrows in Figures~\ref{over1} and \ref{mfs_apr15}. After looking at movies of the slit-jaw, AIA and spectral images, we selected what looked like single events and investigated the line profile evolution in more detail. The selected events are circled in the `width' image, Figure~\ref{iw1}(b). Red circles surround the two events, described in the next sections, that were particularly interesting
  because of their evolution in the slit-jaw images. The events were all oriented so that their length in the east-west direction was longer than their width in the north-south direction. Thus the slit crossed a small part of a larger event as seen in the slit-jaw images. 
 The line profiles of the events circled in blue are shown in the online appendix. The extent of the broad  profiles along the slit is generally only a few pixels, although we show one event (event 5) in the  appendix that extends about 4\arcsec\ (24 pixels) along the slit. 

\subsubsection{Transition region 'jet'}
\begin{figure}
%\centering
\includegraphics[width=8.5cm,clip=]{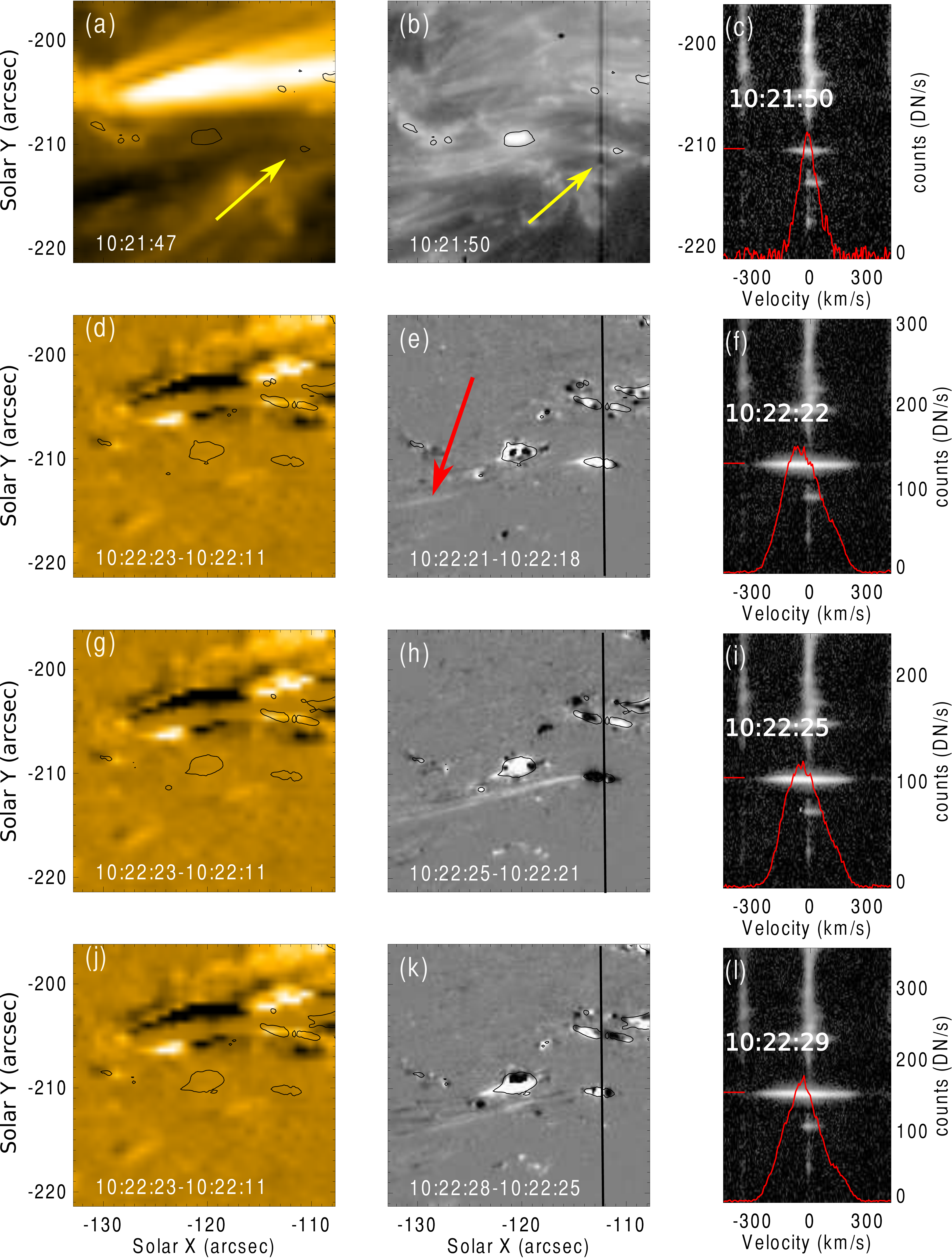}
\caption{Observations of event 1. The top row shows coaligned (a) AIA
171~\AA, (b) IRIS 1400~\AA\ slit-jaw, and (c) IRIS spectral images around \SiIV\ 1402~\AA. The spectral image is overplotted with the line profile from the position marked with a red bar on the left. The other rows show 171 and 1400~\AA\ difference images and spectra at selected times. The site of the event is indicated with yellow arrows in the top row.  The red arrow in (e) points to a bright thread, mentioned in the text. The 171 and 1400~\AA\ images are overplotted with the 1400~\AA\ contours at 200~DN~s$^{-1}$ from the image taken at the later of the two times given on the 1400~\AA\ difference image. The evolution of the event can also be seen in the movie, {\it anim1}. } \label{jet1}
\end{figure}

This event, labelled 1, produced the broadest lines of the sequence and occurred at the footpoint of what appeared to be a jet in the 1400~\AA\ slit-jaw images, on the edge of the positive polarity sunspot (Figure~\ref{mfs_apr15}(a)).The development of the event is shown in  Figure~\ref{jet1} and in the movie {\it anim1}. The top images in Figure~\ref{jet1} show the site, marked by arrows,  before
 the onset of the event.  In the following rows, we show difference images that highlight the key structure changes. The corresponding intensity images and the positions of the 1400~\AA\ brightenings with respect to the magnetic field are shown in the associated movie. 
 The first difference images show that the site brightened at 1400~\AA\ but not in 171~\AA, indicating that there may be cold material absorbing transition region emission with wavelength shortward of the Lyman limit.
 The  1400~\AA\ brightening is  about 3.5\arcsec\ in the east-west direction, compared with 1\arcsec\ along the slit. Since the slit is 0.33~\arcsec\ wide, it only covers about a tenth of the brightening. As seen in (e), the slit  cut across the centre of the 1400~\AA\ brightening. The brightening coincided with an increase in the line width to 110~\kms\ (Figure~\ref{jet1}(f)).
  
A jet-like structure leading directly  from the bright site appears in the next difference image (Figure~\ref{jet1}(h)). It is hard to measure the extent of the jet because it blends with a similar narrow 1400~\AA\ thread  seen in the earlier difference image (red arrow in (e)). Assuming that the jet ends before this thread,  its length is 9\arcsec. Given the time between slit-jaw images, 3.5~s, the speed of the jet would be about 2000~\kms\ which is extremely fast for transition region plasma. We suggest two possible explanations. The first is that the thread was heated to a higher temperature and cooled rapidly along its length thus giving the impression of a fast jet. The second is that the thread is heated by high energy particles generated in the reconnection region that precipitated along a thread-like structure while propagating away from the reconnection site. 
%We note that the transition region brightness of the main brightening in the slit-jaw images decreases when the 'jet' appears, so the appearance of the 'jet' may be related to a release of energy from the site. 
The only evidence for earlier heating is the flare at 9:53~UT. This may be related but because the flare emission is clearly separated from the thread, we think that the second explanation is most likely, in which case the observed brightening and broad line profiles at the footpoint were due to reconnection. The geometry of the jet and flows are quite a puzzle because the broad line profiles extend out to 300~\kms\ which is close to the Alfv\'en velocity, so we expect that the reconnection flow would be toward and away from the observer; however the jet is seen perpendicular to the line-of-sight. 
Some more comments on the jet's appearance and an energy estimate for the particles is given in the discussion section.
 %Without more information it is not possible to determine the cause of the jet. 
%A similar bright thread was seen near one of the other sites  discussed here (see movie, {\it anim9}) but its source was unfortunately not observed by the spectrometer.  
 
\begin{figure}
%\centering
\includegraphics[width=8.5cm,clip=]{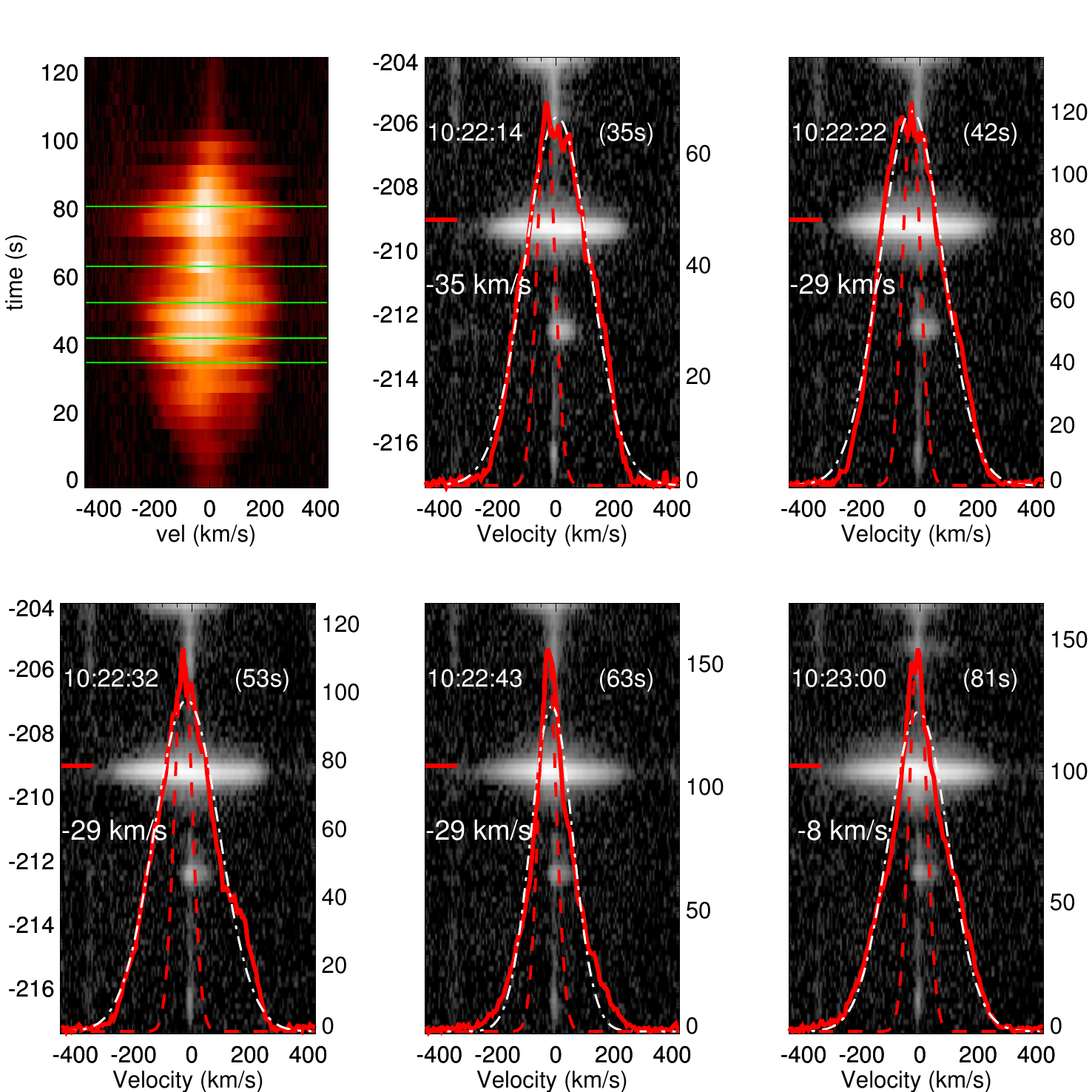}
\caption{Time evolution of the line profiles during event 1. The top left-hand frame shows the time evolution of the profile at the position of the event. Green horizontal lines indicate the times of the spectra in the other frames of the figure. The  spectral images show the event structure along the slit (Solar-Y is on the left axis). The profiles are the average over 5 pixels (0.835\arcsec) at the position of the red horizontal bar on the left. Count rates (DN~s$^{-1}$) are given on the right axis. The narrow, red-dashed profile is a Gaussian with width 30~\kms, centred at the peak intensity of the line. The velocity shift of the line maximum is given on the left of the profile. The white dot-dashed line is the best-fit Gaussian.} \label{pr1}
\end{figure}

The line profile during the evolution of this event is shown in Figure~\ref{pr1}. Broad wings are seen for 90~s, and extend out to $\pm300$~\kms.  During the first 30~s of the event, the core and wings brighten simultaneously and the profile is close to Gaussian (white dot-dashed line) with a width of  about 110~\kms. After about 10:22:32~UT, the core intensity relative to the wings increases and it is no longer possible to fit both the core and wing with a single Gaussian. The core component is well represented by the red-dashed line showing a Gaussian with width 30~\kms.  Like the turbulent events first reported by \citet{BB83}, the profiles are symmetric and there is no offset of the red and blue components along the slit. 
 
\subsubsection{Loop reconfiguration}
\begin{figure}
%\centering
\includegraphics[width=8.5cm,clip=]{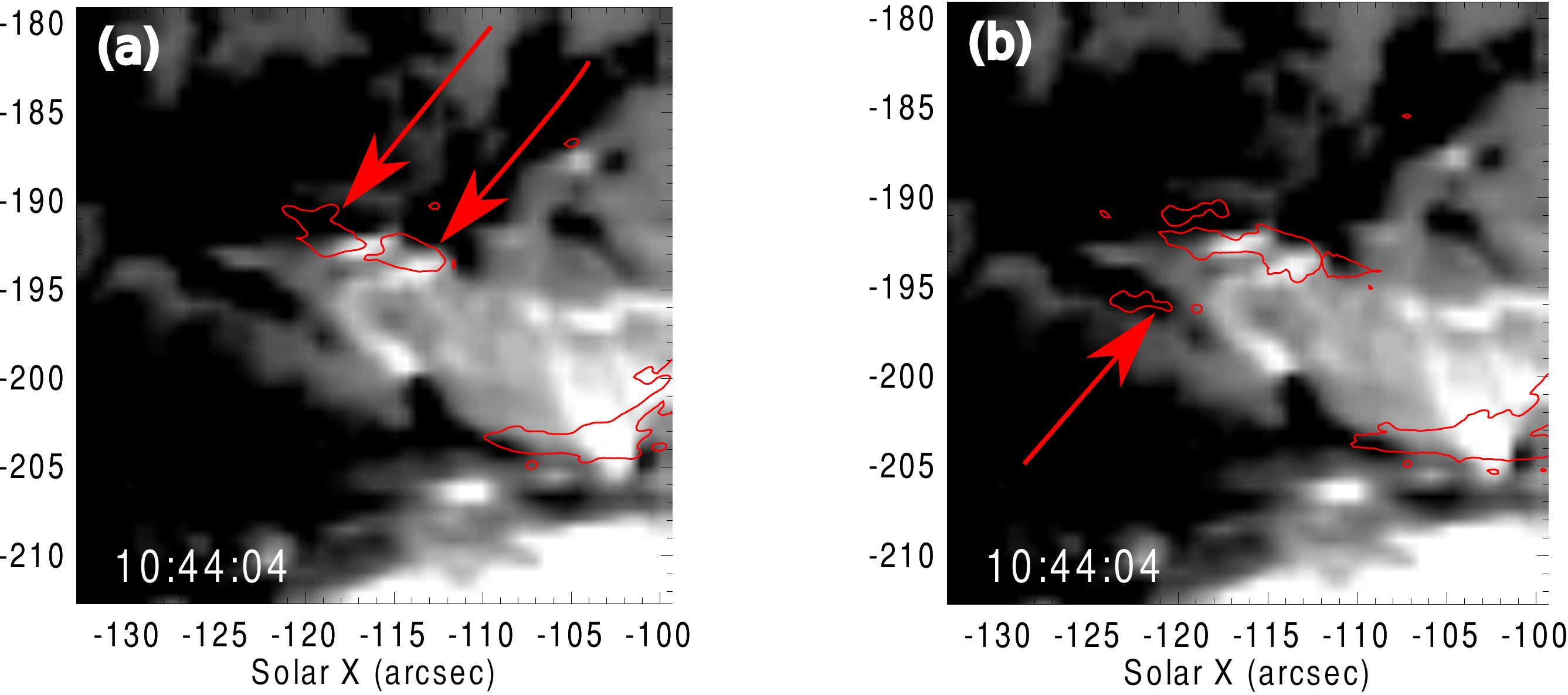}
\caption{The relation between magnetic field and 1400~\AA\ during event 3. The red contours are the 1400~\AA\ intensity at 200~\DNs. The  arrows in (a) point to 1400~\AA\ sites above adjacent positive and negative flux. The arrow in (b) points to the footprint of the new 171~\AA\ loop.} \label{mfs2b}
\end{figure}

\begin{figure}
%\centering
\includegraphics[width=8.5cm,clip=]{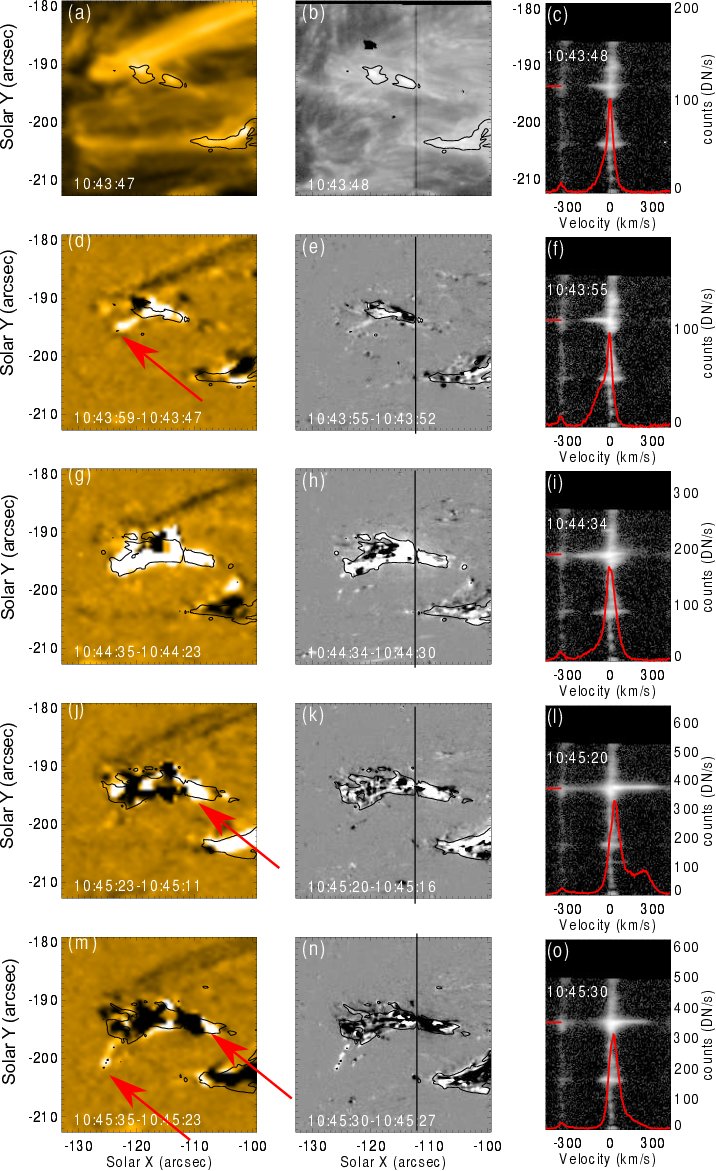}
\caption{Evolution  of event 3. For a description of the images see Figure~\ref{jet1}. The  arrow in (d) points to a new 171~\AA\ loop. The arrow in (j) points to a brightening west of the slit that may be related to redwing enhancements in the spectra. The arrows in (m) point to plasma outflows with a plane-of-sky velocity about 250~\kms. The evolution of the event can also be seen in the movie, {\it anim3}} \label{ev2b}
\end{figure}

\begin{figure}
%\centering
\includegraphics[width=8.5cm,clip=]{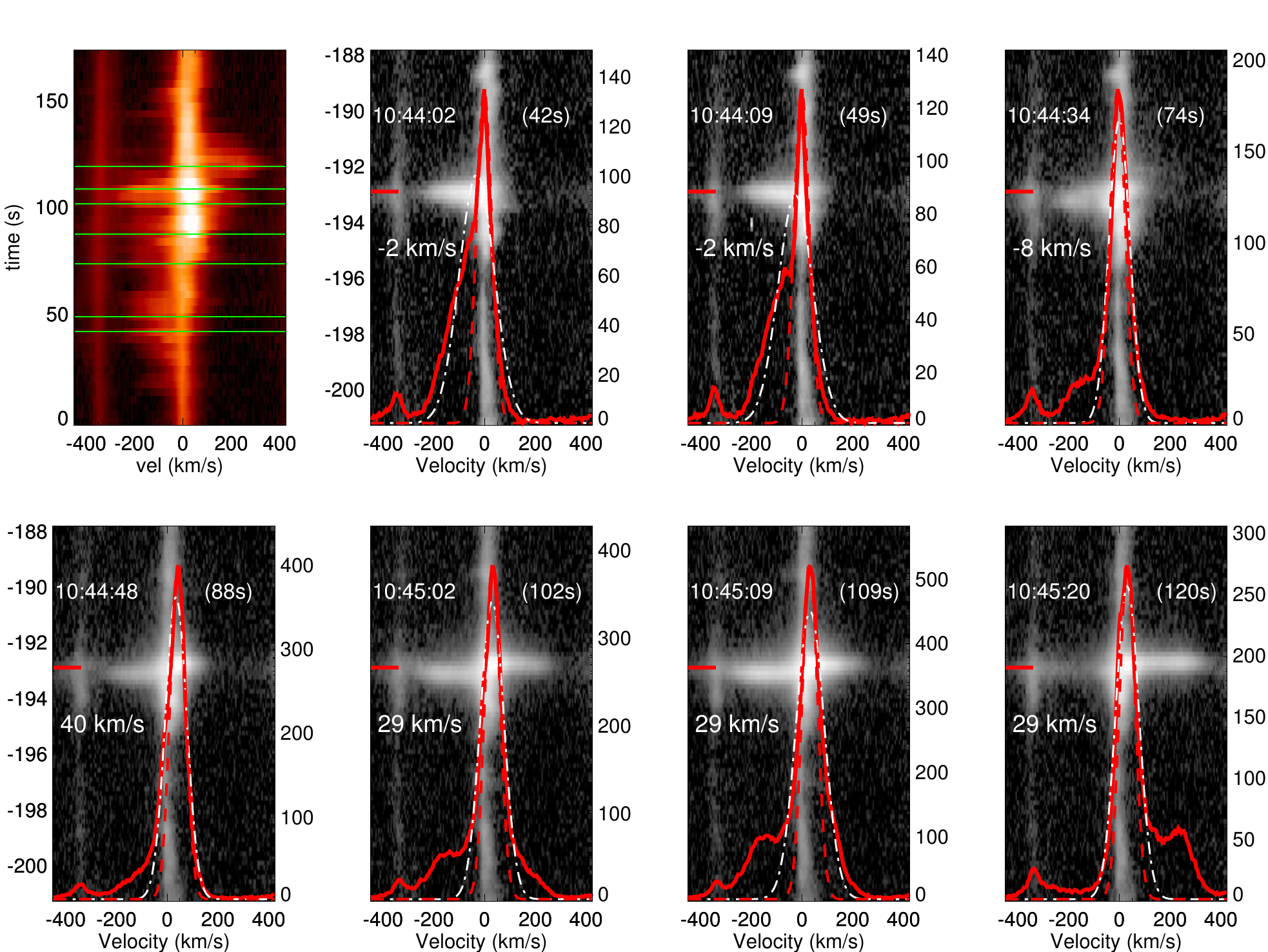}
\caption{Time evolution of the line profiles during event 3. For description of the images see  Figure~\ref{pr1}} \label{pr3}
\end{figure}

The second event, labelled 3, occurred on the edge of the negative sunspot (Figure~\ref{mfs_apr15}(b)). The event was significantly larger than event 1, and its EUV emission was not obscured by intervening cold material so that it was visible  in both the 171 and 1400~\AA\ images. 
%These images show brightening over an area $20\arcsec\times5\arcsec$. 
Before the event there were just two small elongated 1400~\AA\ emission sites on the edge of the negative field  (red arrows in Figure~\ref{mfs2b}(a), Figure~\ref{ev2b}(b)). At event onset a finger of 171~\AA\ emission appeared connecting these sites to negative flux in the south (red arrow in Figure~\ref{mfs2b}(b), Figure~\ref{ev2b}(d)), suggesting that reconfiguration had occurred. The evolution is best seen in the movie, {\it anim3}, of this event. 
%Figures~\ref{ev2b}(a)-(h), and \ref{mfs2b} also show how the 1400 and 171~\AA\ emission changes the connection at the start of the event. A red arrow on the first of the running difference images points to the new 171~\AA\ emission that as shown in Figure~\ref{mfs2b}, connects to a small concentration of negative flux just outside the main spot. 
%The movie shows three strong 171~\AA\ brightenings during the course of the event near the slit. The first two were on the east of the slit and the third, at the time of the last images in Figure~\ref{ev2b}, was on the west. 
%The first brightening was associated with extensive blue-wing enhancements,  the second occurred at the same site and was associated with both blue and red wing enhancements.   During the third, on the west, red-wing enhancements were seen. 
The profile evolution is shown in Figure~\ref{pr3}. Initially the blue wing dominates. Subsequently the line core brightens by a factor three. Both blue and red-wing enhancements are seen with the red wing north of the blue by 0.33\arcsec\  (2 pixels).  Between 10:45:23 and 10:45:11~UT (Figure~\ref{ev2b}(j)) a bright knot of emission appears on the west of the slit. It seems to be related to the red-wing enhancements seen at that time. At the end of the sequence, blobs are seen in the 171 and 1400~\AA\ images moving both to the south and west with a plane-of-sky velocity 
250~\kms\ (Figure~\ref{ev2b}(m)).

 The sequence of wing enhancements could be interpreted as the ejection and retraction of plasma \citep{Huang14},
 %at the position of the slit. 
and the line-core brightening when the red-wing enhancements appear
% is well-fitted with a Gaussian of width 30~\kms. This 
could be explained as plasma compression at the site of the returning plasma. However, as we point out in the discussion section, both the red and blue-wing enhancements could also be due to reconnection outflows, especially as the images show clear evidence for reconfiguration and oppositely directed plane-of-sky flows at the end of the sequence. 
 %sit-and-stare limitations, it is not possible to determine if the red wings enhancements are due to retracting plasma or reconnection outflows. 

\subsection{2014 May 03 observations of AR 12049}
\begin{figure}
%\centering
\includegraphics[width=7cm,clip=]{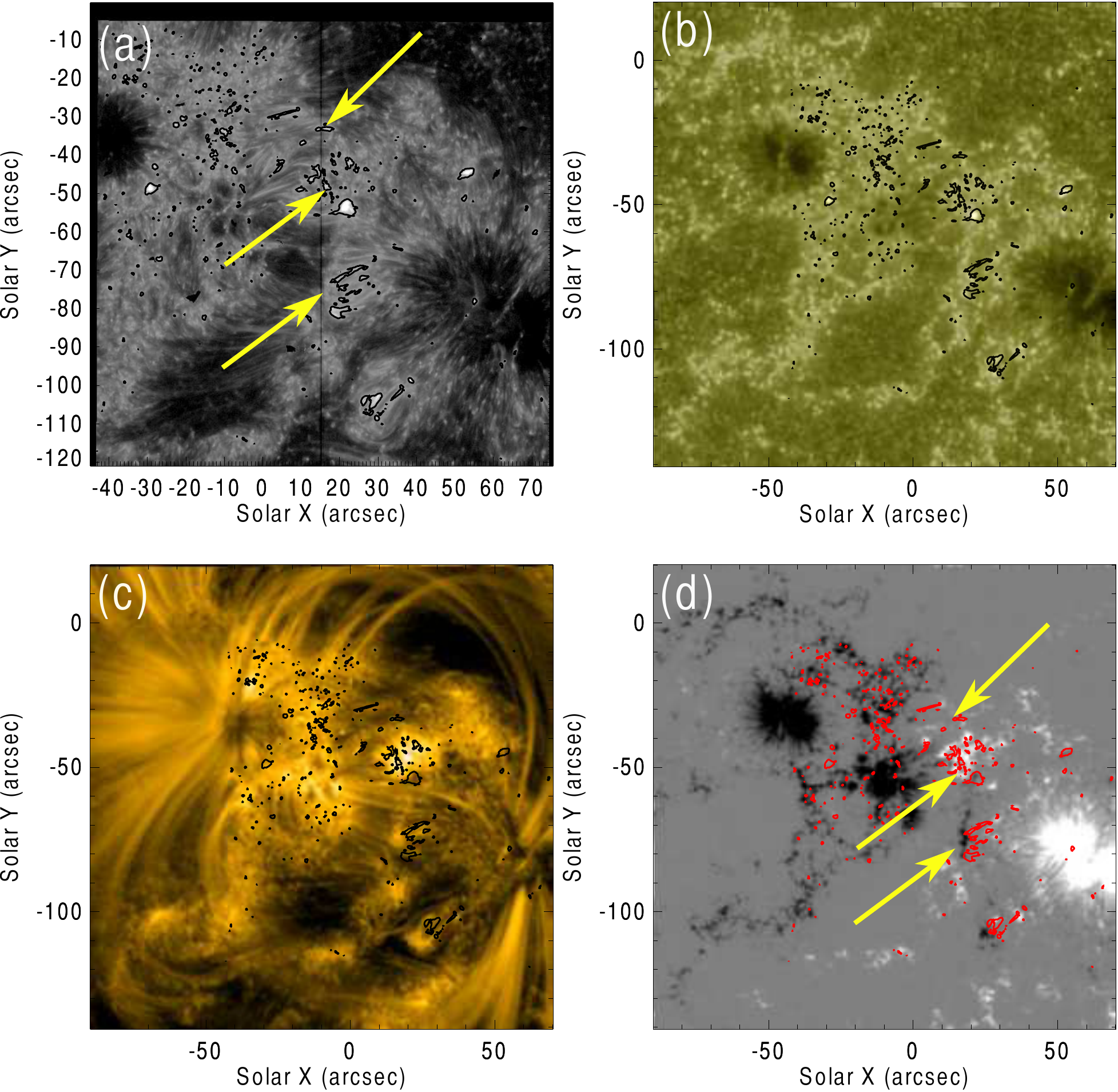}
\caption{AR 12049 on 2014 May 03 13:35:18~UT, site of events 6-15:
(a) IRIS 1400~\AA\ slit jaw; (b) AIA 1600~\AA;
(c) AIA 171~\AA; (d) HMI line-of-sight magnetic field. The black vertical line in
(a) is the IRIS spectrometer slit. All images are overplotted with the
same 1400~\AA\ contours. In (a) and (d) the yellow arrow points to the three main event sites.} \label{over2}
\end{figure}

The second active region,  AR12049, shown in Figure~\ref{over2} was quieter. There was a large leading positive spot separated by about 50\arcsec\ from two trailing negative spots. The previous significant flare from this region had been a  C1.5 two and a half days earlier. The IRIS spectrometer slit was orientated north-south between the positive and negative polarity spots. 
IRIS observations were taken over a period of 3 hours, with an exposure time of  4~s, and a time cadence of 5.6~s for the spectra and 11~s for the 1400~\AA\ slit-jaw images.

\begin{figure}
\includegraphics[width=8.5cm,clip=]{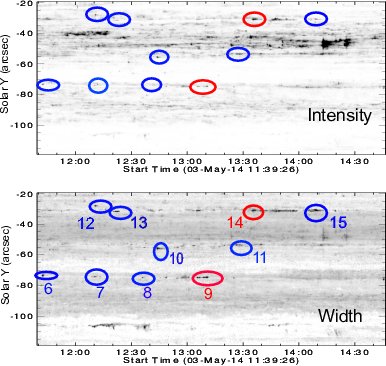}
\caption{\SiIV\ 1402~\AA\ (a) intensity and (b) width during three hours of sit-and-stare.
The maximum values are 1000 DN~s$^{-1}$ and 100~\kms, respectively. 
} \label{iw2}
\end{figure}

In Figure~\ref{iw2} we show the \SiIV\ 1402~\AA\ line intensities and widths. There were many small, apparently isolated, events. The ones characterised by enhanced line wings were mostly at three positions along the slit. They coincide with the regions of mixed-polarity field observed by the spectrometer.  As examples we show in the next section  events, circled in red, from two of the sites. Line profiles from the events inside blue circles are shown in the appendix. The part of the slit covering strong positive field (-50\arcsec $<$ Solar Y $<$ -35\arcsec)  detected many brightenings without broad wings. The part of the slit covering weak field (-70\arcsec $<$ Solar Y $<$ -60\arcsec) showed neither intensity variations nor line broadening.

\begin{figure}
%\centering
\includegraphics[width=8.5cm,clip=]{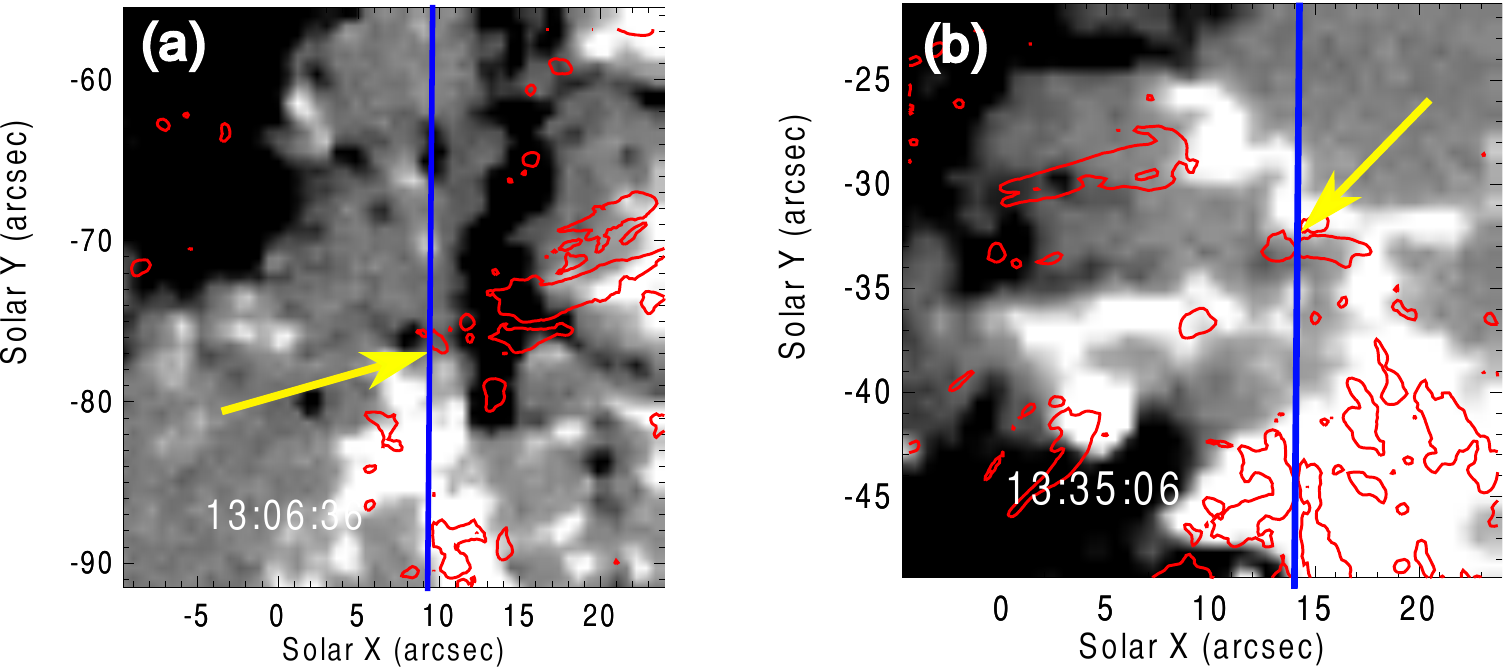}
\caption{Magnetic field at the sites of (a) event 9 and (b) event 14. The red contours are the 1400~\AA\ intensity at 100~\DNs.
} \label{mfs2}
\end{figure}

\subsubsection{Multiple small jets}
The  site with mixed-polarity field, shown in Figure~\ref{mfs2}(a), produced a series of small events in the 3 hours of observation. Although we did not measure the cancellation of photospheric flux, the magnetic configuration was favourable for reconnection in the atmosphere above.  Here we show and discuss the evolution of the last and largest event, event 9, from this site. The line profiles of events marked 6, 7, and 8 are shown in the appendix. 

\begin{figure}
%\centering
\includegraphics[width=8.5cm,clip=]{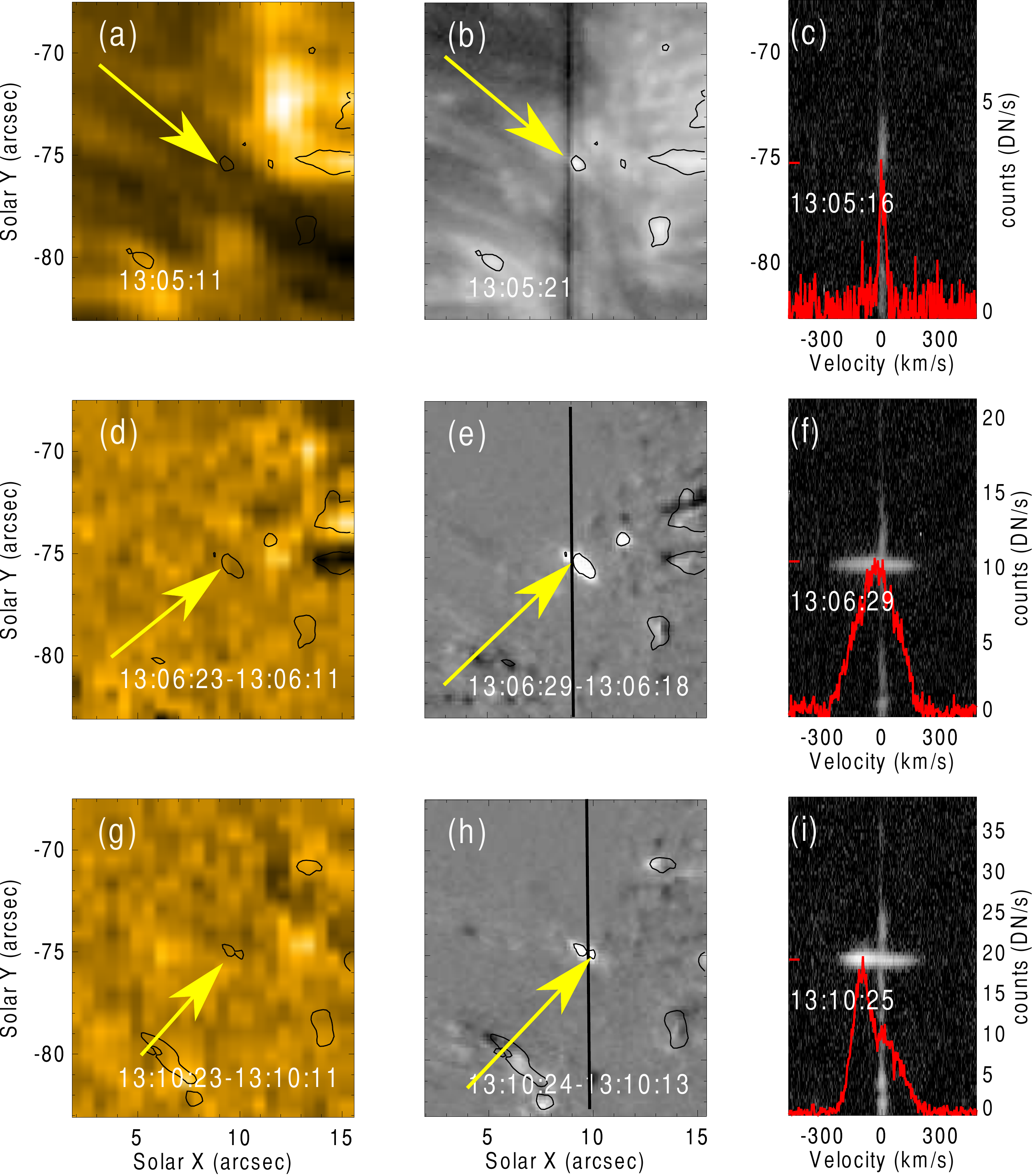}
\caption{Observations of event 9.  For a description of the images see the caption of Figure~\ref{jet1}. The contours are at 1400~\AA\ 100~\DNs. The event evolution is shown in the movie, {\it anim9}}\label{rc9}
\end{figure}

\begin{figure}
%\centering
\includegraphics[width=8.5cm,clip=]{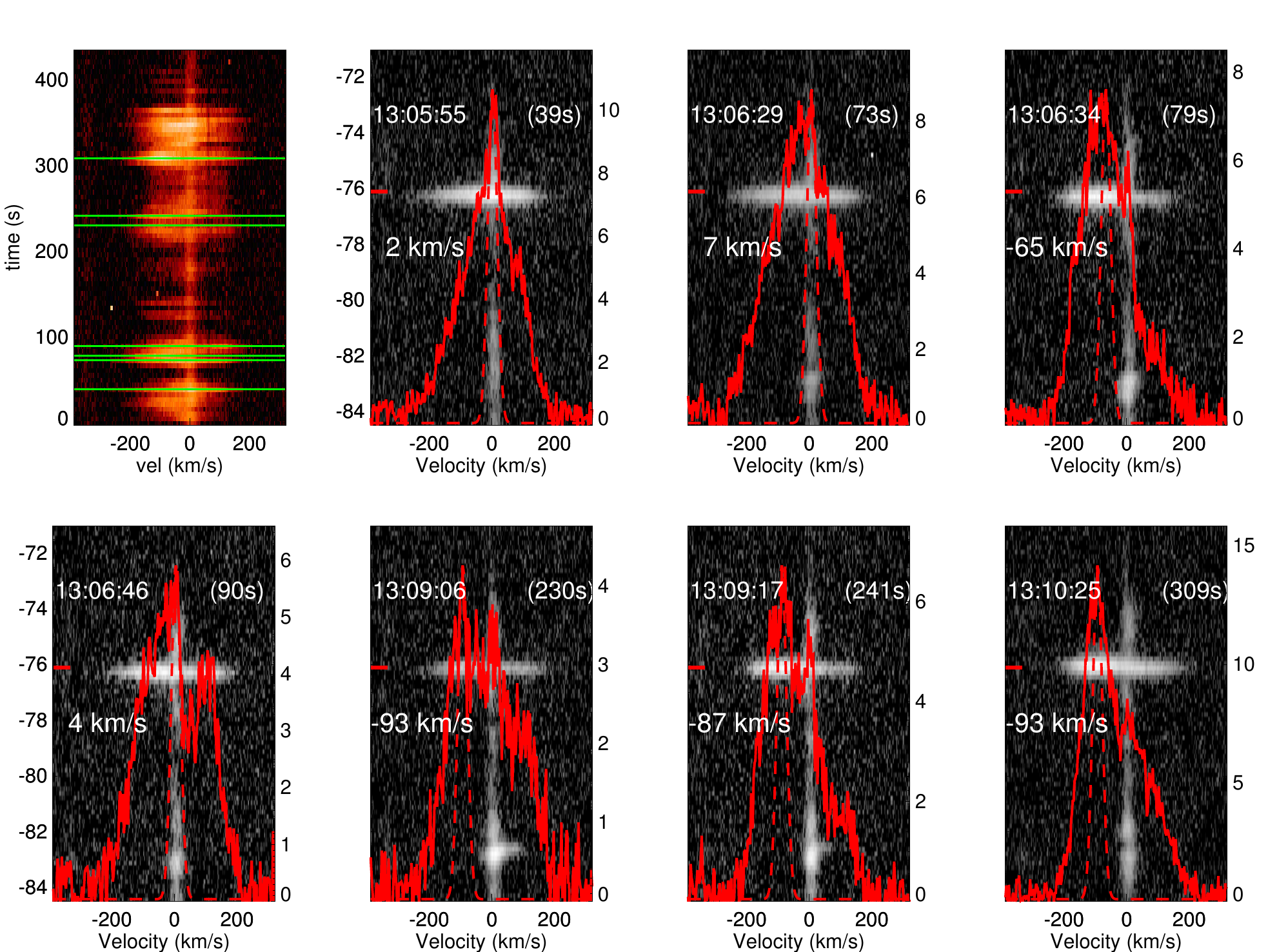}
\caption{Time evolution of the line profiles during event 9. See the caption of Figure~\ref{pr1} for a description of the images.} \label{pr9}
\end{figure}

Selected images and spectra of event 9 are shown in Figure~\ref{rc9}. The full event is shown in the movie, {\it anim9}, and the evolution of the line profile in Figure~\ref{pr9}. It was not one continuous event but consisted of  four events with lifetime of about 50~s and two very short ones with a lifetime about 10~s.  During most of the events the slit is on the eastern edge of the brightening and the profile was distinctly non-Gaussian with a  dominant blue-wing enhancement, suggesting that the slit is crossing the upward propagating part of the reconnection outflow.   The last profile in Figures~\ref{rc9} and \ref{pr9} has an intense blue peak around 100~\kms, tailing off to a broad, low-intensity red wing extending to  200~\kms, as though the up- and downflows were continuous. This is unusual because normally when both blue and red wings are present, they are separated by a bright core.

The profiles from the  other events from this site are shown in the appendix, Figures~A1 to A11. They all show stronger blue than red wings. This is consistent with  outward directed jets being on the east of the brightening, where we see a bigger extension in the images. 
We note that in the EUV images, the site was initially dark  and brightenings were not seen indicating that neutral material in the upper atmosphere obscured the emission at these wavelengths.

\subsubsection{EUV brightening}
\begin{figure}
%\centering
\includegraphics[width=8.5cm,clip=]{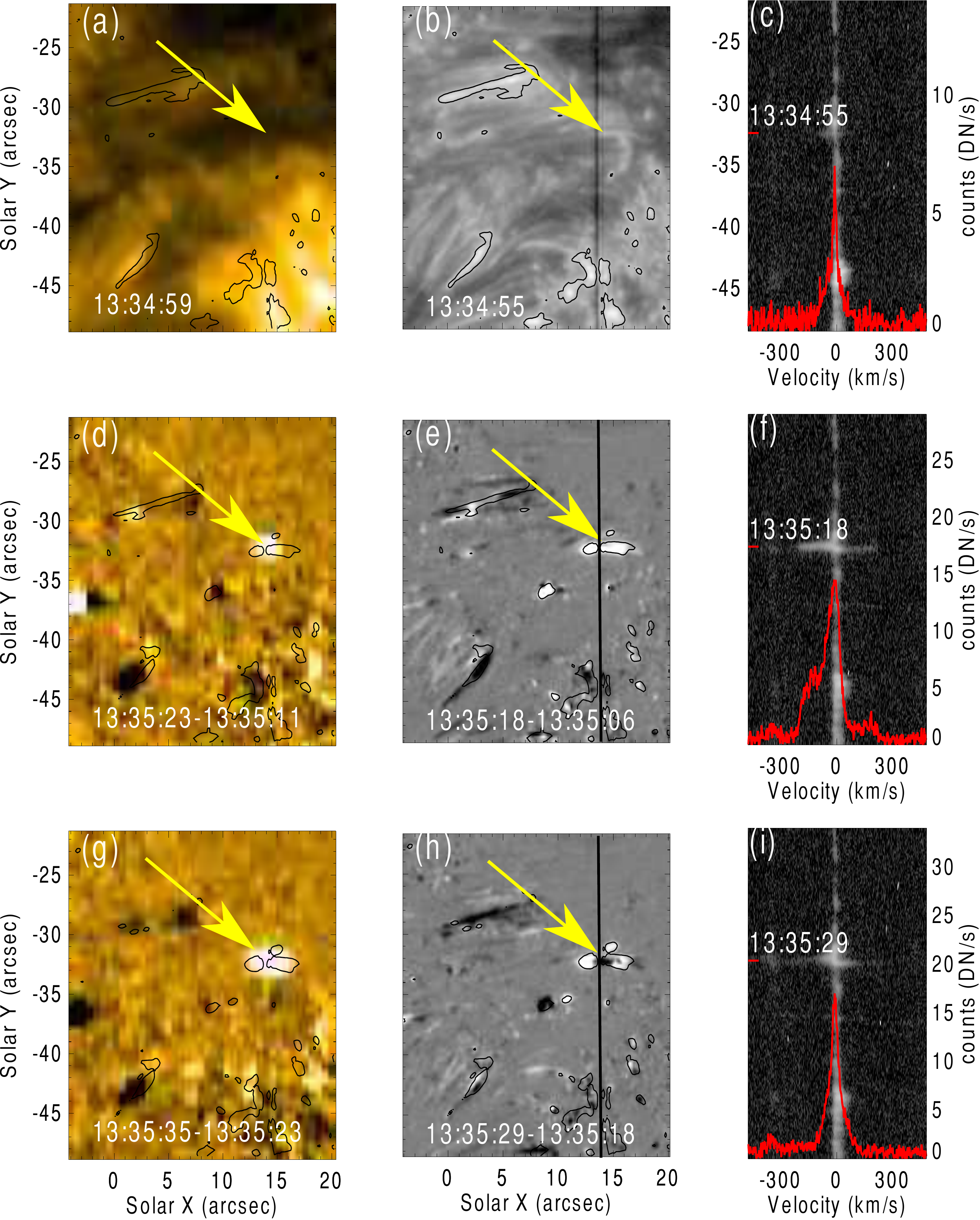}
\caption{Observations of event 14. See caption of Figure\ref{jet1} for a description of the images. The event evolution is shown in the movie, {\it anim14}} \label{rc14}
\end{figure}

This event is noteworthy because in spite of being inside a dark EUV lane, it was seen in the EUV (Figure~\ref{rc14} and {\it anim14}). It occurred on the edge of  positive polarity flux near the north of the slit (Fig~\ref{mfs2}(b)). The event was relatively small  with a lifetime of about 60~s. It started with a clear blue-wing  enhancement and brightening at 1400 and 171~\AA.
Then, as shown in Figure~\ref{pr9}, the line core increased in intensity and the enhanced wings became more symmetric.  
Although the 11~s cadence was a bit slow to pick up movement,  the slit-jaw running-difference images did show increased emission in both the east and west directions (Figures~\ref{rc14}(h) and {\it anim14}) which could be caused by plasma motion away from the event centre.

\begin{figure}
%\centering
\includegraphics[width=7cm,clip=]{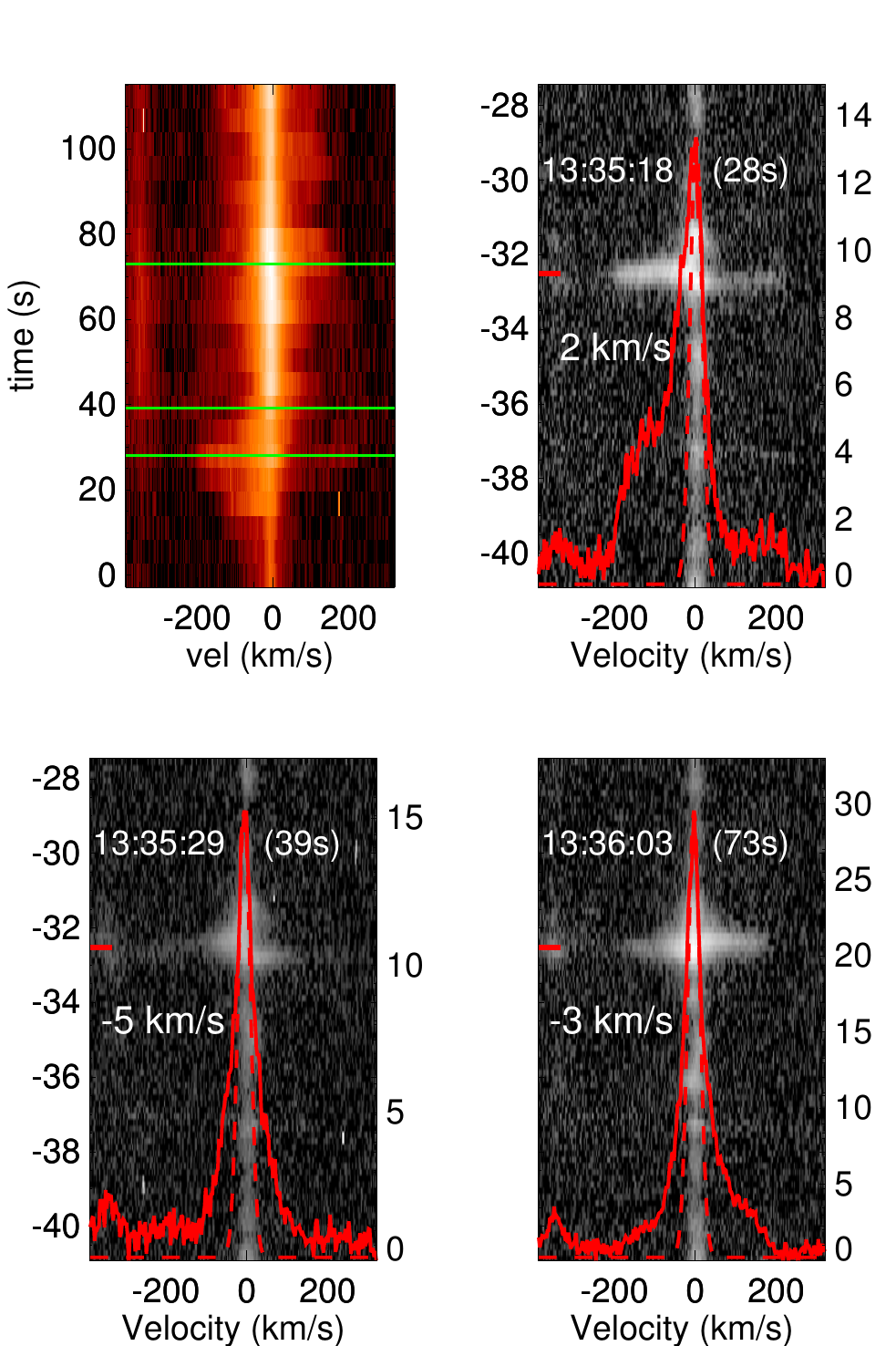}
\caption{Time evolution of the line profiles during event 14. See caption of Figure~\ref{pr1} for a description of the images.} \label{pr14}
\end{figure}

\section{Discussion}
Altogether we show the evolution of \SiIV\ lines profiles from 15 events, from two active regions, with the aim of obtaining an overview of typical explosive event behaviour. The evolution of two events from each active region is described in the main paper and profiles for the other events are shown in the appendix.
We have used sit-and-stare observations and show the full profiles to emphasise the flow structure along a single line-of-sight rather than the energetics. 

%In high spectral and spatial resolution observations of transition region lines from the Sun, broad non-Gaussian profiles are very common. 
The properties of explosive events have been frequently reported but there is still no clear consensus on the profile evolution or their interpretation. Because the sites of line broadening are found at the boundary of strong field regions or close to the neutral line of magnetic bipoles, they are thought to result from magnetic reconnection. 
In this study we found all sites were on the edge of strong field regions, and at half the sites neighbouring opposite polarity field was also present. 
The slit-jaw images always showed brightening at the site and, in several  cases, elongation of the emission structures in opposite directions away from the event centre (events 3 and 14). These motions were perpendicular to and had a shorter lifetime ($<$10~s) than the observed Doppler flows, so they were not the explosive event flows but showed that there was some movement of plasma across the solar surface as well. 

The most surprising of the slit-jaw observations is the jet-like extension that suddenly appeared connected to explosive event site 1 (Figure~\ref{jet1}). As mentioned in section 2.1.1, the `jet' was about 10 times faster and perpendicular to the explosive event flows. We therefore attribute it to a process other than flows, possibly heating by energetic particles.    An estimate of the appropriate particle energies can be obtained by considering  electrons with energy $\epsilon = 2$~keV, interacting with plasma of density $n=10^{10}$~\cc. The collision rate is about 
%$8\times10^{-6}\epsilon^{-3/2}n\ \text{ln}\Lambda$~s$^{-1} = 0.8 $~s$^{-1}$, 
$4\times10^{-6}\epsilon^{-3/2}n\ \textrm{ln}\Lambda$~s$^{-1} = 4.5$~s$^{-1}$, 
where ln$\Lambda$ is the coulomb logarithm which has a value of about 10 \citep{nrl04}. Such particles will travel roughly 6~Mm equivalent to 8\arcsec\ in their 0.2~s lifetime. Thus if particles can be accelerated to about 2~keV in the source region which is quite feasible given the efficiency of particle acceleration,  heating by energetic particles is not an implausible explanation for the thread. However,  it is not easy to understand why only this narrow thread brightened at 1400~\AA. It may be an effect of the coronal (guide) field, or non-uniformities in either the source region or adjacent loops.
Further observations are required to find out how frequently these jet-like structures are seen.

For this study, we note that if it is caused by energetic particles then the \SiIV\ emission is likely from a reconnection site, and the line profile gives information about the velocity and density distribution in the reconnection region.
Previous reconnection models have concentrated on reproducing the highest observed velocities \citep{Detal91,IT99,Retal01}. 
One of the issues not previously addressed is the cause of the low velocity intensity increase observed as bright line cores.  
%About 70\%\ of the investigated quiet Sun events near disk centre produced brightening of  line core \citep{Ning04}. 
In this study,  most profiles have a narrow core with low intensity wings extending out to about 200~\kms (e.g. the profiles in events  2, 3, 4, 5, 10, 11, 13, 14, 15). In most events core brightening was seen soon after event onset (events 1, 2, 3, 5, 10, 11, 12, 13, 14). In fact, the only events that did not show significant core brightening were events $6-9$. These were all from the same site and were the only ones with the line peak significantly shifted 
 ($>50$~\kms) to the blue, so the reason may be due to the geometry of the event and  how the slit crossed the site. As seen in 
 Figure~\ref{rc9}, the slit was on the edge of event 9. 
 Events 10 and 11 had an extended red but very little blue wing  throughout their evolution.  Both were from the same site so this again suggests that the position of the slit with respect to the site is important.    
 In two events the red wing develops at the end of the event (3, 5). Both these events were large complex events, making them difficult to interpret. 
%Only one event, event 3, had a distinct red-wing component for more than one exposure.
 We note that in none of the events did the site of the explosive event move along the slit (in the north-south direction) suggesting that the process causing the shifts occurs at a stationary site. In addition the core and wing emission coincide along the slit, which suggests that they are produced by the same structure not by heated chromospheric plasma below the jet.
 %fast plasma ejections were seen at 1400~\AA. The line-of-sight velocities are typically 200~\kms, so the plasma should move about 3 (9) pixels a 4~s (11~s) exposure.   Motion is however extremely difficult to see in the slit-jaw images because there is so much small-scale activity, and therefore we cannot rule out flows. What is clear is that the site of the explosive event does not move along the slit (in the north-south direction), suggesting that the process causing the shifts occurs at a stationary site.  
 
We now assume that the basic cause is reconnection and test whether  the line profiles can, at least qualitatively, be reproduced by reconnection models. Previous  models of  fast reconnection in the transition region \citep{IT99,Retal01} have used anomalous resistivity at a single site to generate pairs of slow-mode shocks \citep{P64}. Recently it has become clear that current sheets are unstable to plasmoid formation \citep{Loureiro07, Bhatt09, Huang13} with a linear growth rate that scales with the Lundquist number, $S$, as  $S^{1/4}$ and a nonlinear reconnection rate that is independent of $S$ \citep{Bhatt09,Huang10,Uzdensky10} in comparison to the Sweet-Parker rate that scales with $S^{-1/2}$. So in high Lundquist number plasmas such as the solar transition region where $S\sim10^{10}$, the short-lived linear regime of the plasmoid instability can witness  fast, exponential growth leading to a nonlinear regime in which multiple islands of dense, low-velocity plasma can be produced.  Each island has an o-point in its centre and a pair of x-points at its sides where plasma is accelerated. 
 The formation and interaction of multiple islands in the current sheet leads to velocity distributions \citep{Loureiro12, Uzdensky10, Huang12} that may produce bright cores and enhanced wings. 

\begin{figure}
%\centering
\includegraphics[width=10cm]{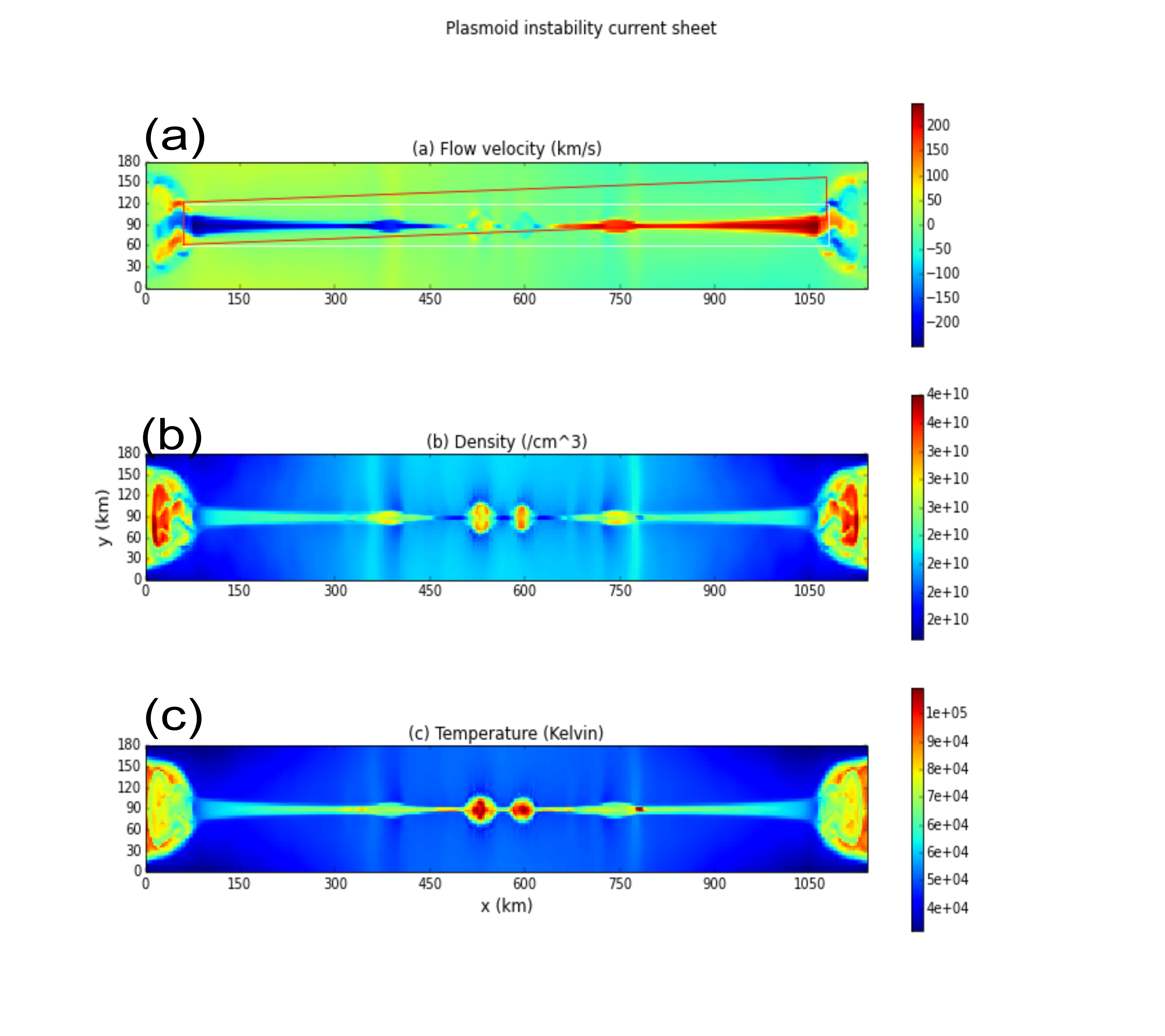}
\caption{Simulation of reconnecting current sheet: (a) velocity (b) density (c) temperature. The white box in (a) shows the region that produced profiles Figure~\ref{sim_pr}(b)-(d), and the red box the region that produced (e)-(g).  The integration was along the long axis of the boxes.}\label{sim}
\end{figure}

Results from two-dimensional (2D) simulations of reconnection along a Harris sheet at Lundquist number $10^5$ are shown in Figure~\ref{sim}. The simulations have been done with the Athena code \citep{Stone08}, using the double current-sheet setup described in \citet{Hawley95}, and a grid with 200 cells along the sheet and 400 across it. The setup did not include a guide field. Reconnection along the current sheet is triggered with a small magnetic perturbation along the current sheet, as described in \citet{Guo14}. Small, random velocity perturbations throughout the domain are used to initiate the instability. The asymptotic values are plasma beta, $\beta = 0.1 $, inflow number density $n=10^{10}$~\cc, and magnetic field $B=12$~G which  gives an Alfv\'en speed in the inflow region, $V_A = 250$~\kms\ and temperature, $T=2\times10^5$~K. Our simulation parameters are a compromise between what the observations require and what is computationally feasible. For example the Lundquist number is much smaller than that of the Sun. This means that the linear instability growth rate is about a factor 10 too slow.    
Ohmic heating is included but radiative cooling and conduction are not which makes the simulated temperatures rather high. Nevertheless the current sheet structure which produces the line-profile characteristics should remain even when  these processes are included. 

\begin{figure}
%\centering
\includegraphics[width=0.95\linewidth]{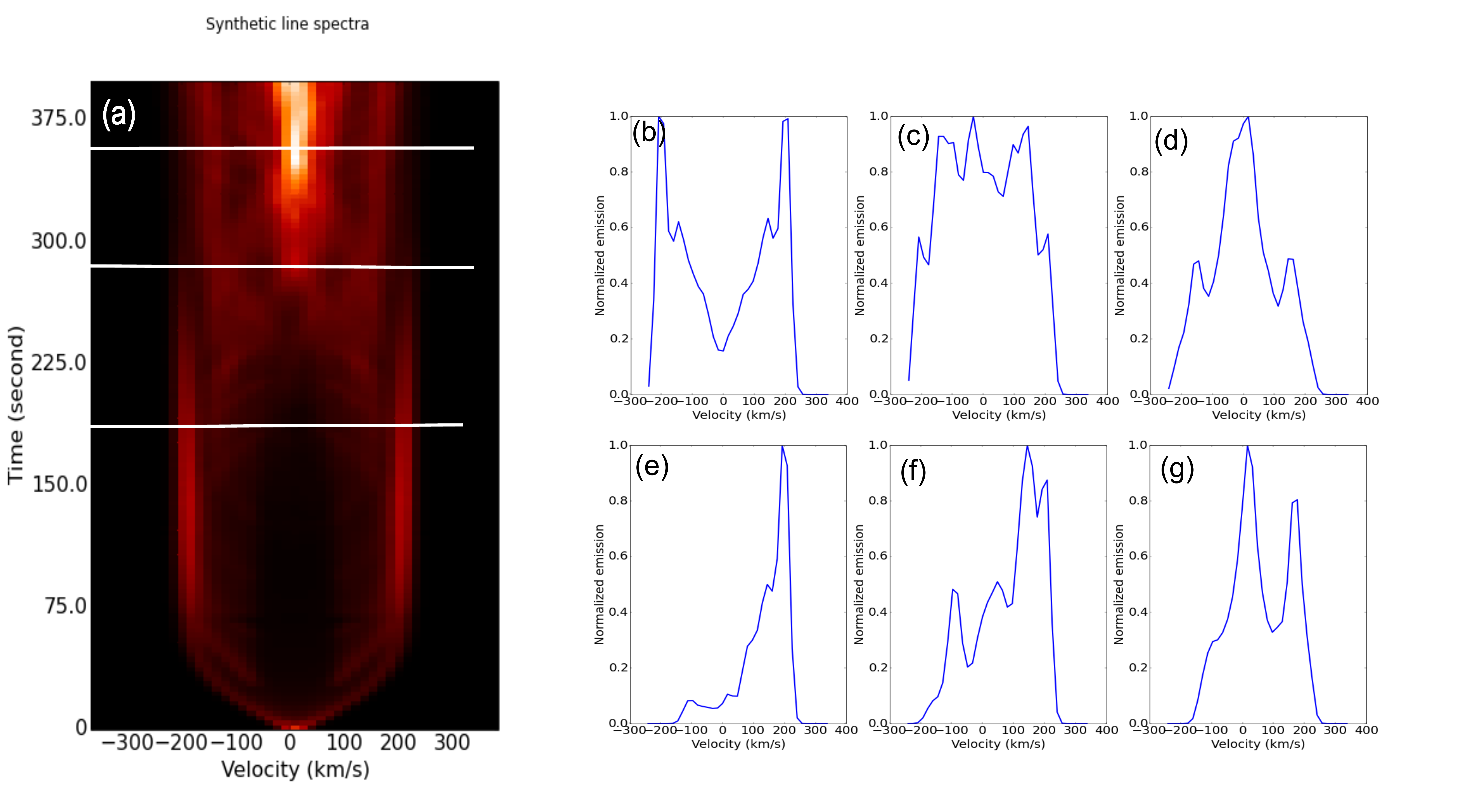}
\caption{Line profiles from simulations of a 2D current sheet with $\beta = 0.1$: (a) time evolution of the line profile of a line formed around $8\times10^4$~K, with line-of-sight along the current sheet (white box in Figure~\ref{sim}(a)). The white horizontal lines indicate the times at which the profiles on the right are taken. The top row shows profiles along the current sheet (white box) and the bottom integration along the line-of-sight  at a slight angle (red box).}\label{sim_pr}
\end{figure}

Figure~\ref{sim} shows the velocity,  number density, and temperature along the current sheet just after the onset of plasmoid formation when there are about 3-4 well-separated  plasmoids along the sheet.  Each of the central high-density, low-velocity islands is separated by bi-directional high-velocity jets. In both directions the maximum plasma velocity is of the order the Alfv\'en speed which in active regions is about $200-300$~\kms.  
Synthetic transition region (e.g. \SiIV) line profiles are shown in Figure~\ref{sim_pr}. For calculating line intensity we assume a top-hat  function that is one between $6\times10^4$ and $1\times10^5$~K and zero everywhere else, and a Gaussian line width based on the plasma temperature.  Initially reconnection produces oppositely directed jets similar to Sweet-Parker reconnection, seen as well-separated red and/or blue wings in the line profile \ref{sim_pr}(b) . Then the current sheet becomes unstable and breaks up into several small plasmoids moving at different velocities reflected in the line profiles as broad, multi-component profiles (\ref{sim_pr}(c)). About 1 min later, the smaller plasmoids merge giving rise to broad profiles with a bright core component (\ref{sim_pr}(d)). Depending on the integration path, profiles with different core-to-wing ratio and asymmetry are obtained. In the bottom row of Figure~\ref{sim_pr}, we show profiles obtained by integration along the red box in Figure~\ref{sim}(a) at the same times. 

\begin{figure}
%\centering
\includegraphics[width=0.95\linewidth]{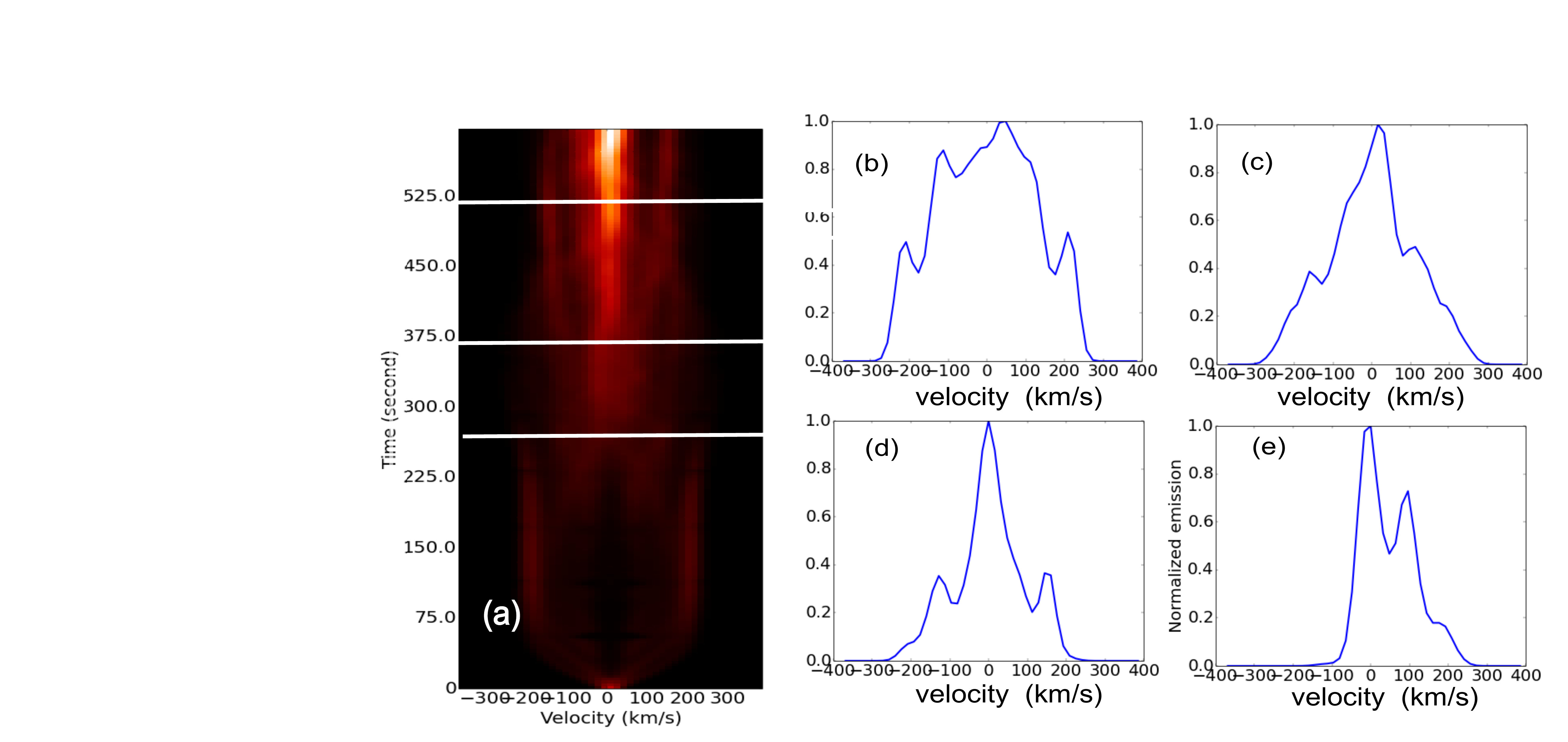}
\caption{Line profiles from simulations of a 2D current sheet  with $\beta = 0.5$: (a) time evolution of the line profile of a line formed around $5\times10^5$~K with line-of-sight  along the current sheet. The white horizontal lines indicate the times at which the profiles (b)-(d) are taken. The profile (e) is taken along a line-of-sight at a small angle to the current sheet at the time of (d).}\label{sim_pr2}
\end{figure}

The island characteristics are affected by plasma beta, grid resolution, and Lundquist number. An increase in any of these parameters leads to smaller islands and and smoother wings. Examples of profiles from a $\beta= 0.5$ run is shown in Figure~\ref{sim_pr2}. Since temperature scales with $\beta$, the temperature of the line shown here is $5\times10^5$~K. Explosive event profiles are seen frequently in lines such as \OVI\ and \NeVIII\ which are formed at these higher temperatures \citep{Winebarger02}.

Although this simulation is not realistic because it is only 2D and does not include radiative losses or conduction, it gives a qualitative idea of how plasmoid formation might affect the line profiles.  They have much better agreement with observations than profiles from  Petschek or Sweet-Parker reconnection  which only have strong wing brightening \citep{IT99,Retal01}, or larger-scale, low-resolution computations that have distinct jet and background core components \citep{Ding11}. 
 

%The configuration of the reconnection site is idealised in the simulation. 
 In reality the magnetic configuration is more complicated and non-uniformities at the reconnection site  are likely to contribute to the line profiles. The point we would like to emphasise is that island formation, and hence the plasmoid instability as a self-contained model, is able to  explain some of the principle features  of the observed profiles in the context of reconnection.  The next step is to investigate raster observations with IRIS to deduce the flow geometry of events, and develop more realistic simulations that include conduction, radiative cooling and a realistic solar atmosphere. Our MHD simulations will need to be augmented by kinetic simulations in order to make connections with the physics of particle acceleration and heating.%Due to the fast evolution and large, compared with the IRIS slit, size of events we opted to analyse sit-and-stare in this study.

\section{Conclusion}
The line profiles at small-scale acceleration sites in active regions consistently show broad, low-intensity red and/or blue wings extending to 200~\kms, and a bright central core. The core with a width of about 30~\kms\ tends to brighten after the first appearance of the wing emission. The core and wing emission are spatially co-incident and neither move significantly during the typically 2-5~min event duration. Thus both components seem to come from the same stationary site. 
It has been long speculated because of their association with cancelling magnetic flux that these line broadenings are caused by small-scale magnetic reconnection.
Previous reconnection models, based on the Petschek mechanism, were able to reproduce the wing but not the core emission. We suggest that the core emission is from high-density, low-velocity magnetic islands that form due to the plasmoid instability along the current sheet under high-Lundquist-number solar conditions.
Our 2D-MHD simulations of magnetic reconnection show the rapid growth of magnetic islands along the current sheet. The islands are separated by fast jets and can explain both the core and wing components of the observed line profiles. We conclude that the IRIS line profiles could result from reconnection via the plasmoid instabilty during small-scale events on the Sun.
 
\begin{acknowledgements}
We would like to thank the referee for constructive comments. This work was supported by the Max-Planck/Princeton Center for Plasma Physics , NSF and NASA.
\end{acknowledgements}

\end{document}